\documentclass[aps,prd,twocolumn,superscriptaddress,amssymb,eqsecnum,showpacs,showkeyes,
secnumarabic,graphics,floatfix,nofootinbib,tightenlines,longbibliography]{revtex4-1}

\usepackage{graphicx}
\usepackage{bm}
\usepackage{cancel}
\usepackage{epsfig}
\usepackage{dcolumn}
\usepackage{amsmath}
\usepackage{hhline}
\usepackage[utf8x]{inputenc}
\usepackage[dvipsnames]{xcolor}
\usepackage{float}
\usepackage[breaklinks=true,colorlinks=true,
linkcolor=blue,urlcolor=blue,citecolor=MidnightBlue,
bookmarks=true,bookmarksopenlevel=2]{hyperref}
\usepackage{mathrsfs}

\def\apj{{Astroph.\@ J.\ }}

\def \beq  {\begin{equation}}
\def \eeq  {\end{equation}}
\def \ber  {\begin{eqnarray}}
\def \eer  {\end{eqnarray}}
\def \omm  {\Omega_{m,0}}

\begin{document}
\newcommand{\newc}{\newcommand}

\newc{\be}{\begin{equation}}
\newc{\ee}{\end{equation}}
\newc{\ba}{\begin{eqnarray}}
\newc{\ea}{\end{eqnarray}}
\newc{\bea}{\begin{eqnarray*}}
\newc{\eea}{\end{eqnarray*}}
\newc{\D}{\partial}
\newc{\ie}{{\it i.e.} }
\newc{\eg}{{\it e.g.} }
\newc{\etc}{{\it etc.} }
\newc{\etal}{{\it et al.}}
\newc{\lcdm}{$\Lambda$CDM }
\newc{\wcdm}{wCDM }
\newc{\lcdmnospace}{$\Lambda$CDM}
\newc{\plcdm}{Planck15/$\Lambda$CDM }
\newc{\plcdmnospace}{Planck15/$\Lambda$CDM}
\newc{\geffz}{$G_{\rm eff}(z)$ }
\newc{\geffznospace}{$G_{\rm eff}(z)$}
\newcommand{\nn}{\nonumber}
\newc{\ra}{\Rightarrow}
\newc{\la}{\label}
\newcommand{\fs}{{\rm{\it f\sigma}}_8}

\title{Consistency of Modified Gravity with a decreasing $G_{\rm eff}(z)$ 
in a $\Lambda$CDM background}
\author{Radouane Gannouji}\email{radouane.gannouji@gmail.com}
\affiliation{Instituto de Fısica, Pontificia Universidad Catolica de Valparaıso, Casilla 4950, 
Valparaıso, Chile}
\author{Lavrentios Kazantzidis}\email{lkazantzi@cc.uoi.gr}
\author{Leandros Perivolaropoulos}\email{leandros@uoi.gr}
\affiliation{Department of Physics, University of Ioannina, GR-45110, Ioannina, Greece}
\author{David Polarski}\email{david.polarski@umontpellier.fr}
\affiliation{Laboratoire Charles Coulomb, Universit\'e Montpellier 2 \& CNRS UMR 5221, F-34095 Montpellier, France}

\date {\today}

\begin{abstract}
Recent analyses \cite{Nesseris:2017vor,Kazantzidis:2018rnb} have indicated that an effective 
Newton's constant $G_{\rm eff}(z)$ decreasing with redshift may relieve the observed tension 
between the Planck15 best fit $\Lambda$CDM cosmological background ({\it i.e.} 
Planck15/$\Lambda$CDM) and the corresponding $\Lambda$CDM background favored by growth 
$f\sigma_8$ and weak lensing data. We investigate the consistency of such a decreasing 
$G_{\rm eff}(z)$ with some viable scalar-tensor models and $f(R)$ theories. We stress that 
$f(R)$ theories generically can not lead to a decreasing $G_{\rm eff}(z)$ for any cosmological 
background. For scalar-tensor models we deduce that in the context of a $\Lambda$CDM 
cosmological background, a decreasing $G_{\rm eff}(z)$ is not consistent with a large 
Brans-Dicke parameter $\omega_{BD,0}$ today. This inconsistency remains and amplifies in the 
presence of a phantom dark energy equation of state parameter ($w < -1$). However it can be 
avoided for $w >-1$. We also find that any modified gravity model with the required decreasing 
$G_{\rm eff}(z)$ and $G_{{\rm eff},0}=G$, would have a characteristic signature in 
its growth index $\gamma$ with $0.61\lesssim \gamma_0\lesssim 0.69$ and large slopes 
$\gamma_0'$, $0.16\lesssim \gamma_0'\lesssim 0.4$, which is a characteristic signature of a 
decreasing (with $z$) $G_{\rm eff}(z)<G$ on small redshifts. This is a substantial departure 
today from the quasi-static behaviour in $\Lambda$CDM with $(\gamma_0,\gamma_0')\approx 
(0.55,-0.02)$.   
\end{abstract}
\maketitle
\section{Introduction}
\label{sec:Introduction}
A wide variety of theories \cite{Copeland:2006wr,Li:2011sd,Capozziello:2011et,Chiba:2012cb,ArmendarizPicon:2000ah,Cai:2015emx,Hu:2007nk,Sahni:2002dx,Bento:2002ps} have been proposed for the 
description of the observed accelerating expansion of the universe. However, the simplest 
model (the \lcdm model \cite{Carroll:2000fy,Peebles:2002gy,Sahni:1999gb}) remains consistent with (almost) all cosmological observations \cite{Sahni:2006pa,Percival:2007yw,Alam:2016hwk,Kowalski:2008ez,Jaffe:2000tx,Izzo:2015vya,Efstathiou:2017rgv,Allen:2007ue,Baxter:2016ziy,Chavez:2016epc,Ade:2015xua,Aghanim:2018eyx}. The best fit 
parameter values of this model have been reported by the Planck mission \cite{Ade:2015xua,Aghanim:2018eyx} 
with extreme accuracy and define the concordance \plcdm model, 
which is consistent with geometric cosmological observations. Such observations include the 
Type Ia Supernova data \cite{Perlmutter:1998np,Riess:1998cb,Suzuki:2011hu,Betoule:2014frx}, 
the Baryon Acoustic Oscillations data \cite{Alam:2016hwk,Percival:2007yw} etc.

However, recent analyses \cite{Bull:2015stt,Sola:2016zeg,Lin:2017bhs,Lin:2017ikq,Sahni:2014ooa} indicate some tension between the 
\plcdm model and some dynamical observations measuring the growth rate of cosmological 
perturbations. Such observations include Weak Lensing data \cite{Hildebrandt:2016iqg,Joudaki:2017zdt,Jee:2015jta,Abbott:2015swa,Kohlinger:2017sxk} and Redshift 
Space Distrortions (RSD) \cite{Nesseris:2017vor,Kazantzidis:2018rnb,Macaulay:2013swa,
Bull:2015stt,Tsujikawa:2015mga,Basilakos:2017rgc}. The robust observable reported by RSD 
surveys is the product
\be 
\fs(z) \equiv f(z)\cdot \sigma(z)= - (1+z) \frac{\sigma_{8,0}}{\delta_0} \delta_m'(z)  
\label{eq:fsigma8}
\ee
where $f\equiv d\ln \delta_m/d\ln a$ describes the growth of 
cosmological matter density perturbations $\delta_m = \delta \rho_m /\rho_m$, and a prime 
stands for the derivative with respect to redshift $z$. 
The quantity $\sigma_8(z)$ is the rms density fluctuation 
on comoving scales corresponding to $8 h^{-1} Mpc$ at redshift $z$ while $\sigma_{8,0}$ refers to the present time value of $\sigma_8(z)$. 

Since 2006 there has been a significant increase of surveys that measure RSDs leading to a 
collection of 63 $\fs$  data points \cite{Kazantzidis:2018rnb}. Despite possible correlations 
among the data points of this dataset its various subsamples considered in the literature 
\cite{Nesseris:2017vor,Basilakos:2017rgc,Basilakos:2016nyg,Macaulay:2013swa} indicate various levels 
of tension between \plcdm parameter values and the parameter values favored by the considered 
$\fs$ subsample. The level of this tension appears to decrease for more recently published 
$\fs$ data \cite{Kazantzidis:2018rnb}. However all considered $\fs$ subsamples seem to 
indicate a reduced growth rate compared to the one expected in the context of \plcdm and GR.

The observed tension \cite{Alam:2015rsa} could be relaxed following one of the following 
methods
\begin{itemize}
\item Modifying the background, \ie considering a smaller value for $\omm$ and/or a smaller 
value for $\sigma_{8,0}$. Other probes, such as the $WMAP$, report lower values for both $\omm$ 
and $\sigma_{8,0}$ \cite{Hinshaw:2012aka}.
\item Considering modified gravity theories  which give a decreasing function of \geffz with 
$z$.
\end{itemize}
In this analysis we investigate the consequences of the second case.

The linear evolution of $\delta_m$ is given by the equation
\be
{\ddot \delta_m} + 2H {\dot \delta_m} - 4\pi G_{\rm eff}\,
\rho \, \delta_m = 0 \la{eq:odedeltat}
\ee
In terms of the redshift, Eq. \eqref{eq:odedeltat} is rewritten as
\be 
\delta_m'' + \left( \frac{ (\ln h^2)' }{2} - \frac{1}{1+z} \right) \delta_m' = 
\frac32 (1+z) h^{-2} {G_{\rm eff}(z,k)\over G}~\Omega_{m,0} \delta_m~,\la{eq:odedeltaz}
\ee
where $\Omega_{m,0}$ is the present relative matter density, $h\equiv \frac{H}{H_0}$ and 
$H_0$ is the Hubble parameter today, $G_{\rm eff}(z,k)$ is 
the effective Newton's constant which for General Relativity (GR) is the usual Newton's 
constant $G$. In general for modified gravity models $G_{\rm eff}$ depends both on the redshift 
$z$ and the scale $k$. 

The central quantity $G_{\rm eff}(z,k)$ comes from a generalization of Poisson's equation
\cite{BEPS00},\cite{EspositoFarese:2000ij}
\be 
\nabla^2 \phi \approx 4 \pi G_{\rm eff} \rho \; \delta_m~, \la{eq:genpoisson}
\ee
while the potential $\phi$ can be read off the perturbed metric in the longitudinal 
(Newtonian) gauge 
\be
ds^2= -(1 + 2 \phi) dt^2 + a^2 (1 - 2\psi) d{\vec{x}}\,^2 \la{eq:newgaugephi}
\ee
Solar system constraints \cite{GPRS06},\cite{Nesseris:2006hp}  imply that 
\be 
\Big\lvert H_0^{-1} \frac{\dot{G}_{{\rm eff},0}}{G} \Big\rvert = 
      \Big\lvert \frac{G'_{{\rm eff},0}}{G} \Big\rvert \lesssim 10^{-3}
\la{eq:geffconstr1}
\ee
whereas they actually leave the second derivative unconstrained since 
\be 
\Big\vert \frac{G''_{{\rm eff},0}}{G} \Big\vert \lesssim 10^{5} 
\la{eq:geffconstr2}
\ee
Thus an interesting question that arises is the following: ``Which modified gravity models
are consistent with $G_{\rm eff}(z)/G<1$ at low z?". A naive answer to this question would 
indicate that any modified theory of gravity can lead to $G_{\rm eff}(z)/G<1$ at low $z$ for 
some appropriate parameter values. In the present analysis we address this question 
and argue that this is not so for at least two important and intensively studied examples, 
the standard, massless scalar tensor gravity model and $f(R)$ models.
More specifically, we address the following questions: 
\begin{enumerate}
\item What is the generic form of $G_{\rm eff}(z)$ at low $z$ for standard scalar tensor 
and $f(R)$ theories when one assumes a \lcdm background expansion? 
\item How do the above answers change for different background expansion rates $H(z)$?
\end{enumerate}

The structure of this paper is the following: In the next section we derive the generic form 
of \geffz for low $z$ for some modified gravity models.
In Sec. \ref{sec:gamma} we consider the behaviour of the growth index in these models.
Finally in Sec. \ref{sec:Conclusion} we summarize and discuss our results.

\section{\geffz in some models}
\label{sec:2}
It is our purpose to investigate whether some modified gravity models allow 
for a decrease of $G_{\rm eff}$ below the usual Newton's constant $G$, its value in GR.

\subsection{$f(R)$ modified gravity models}
The answer is negative \cite{Polarski:2016ieb} for viable $f(R)$ models, see 
e.g. \cite{Hu:2007nk}, \cite{S07}. This can be 
seen immediately from the expression of $G_{\rm eff}$ in these models, viz. 
\be
\frac{G_{\rm eff}(z,k)}{G} = \left( \frac{df}{dR} \right)^{-1}~\left[ 1 +
\frac{\left(\frac{\lambda_c}{\lambda}\right)^2 }{3\left(1 +
\left(\frac{\lambda_c}{\lambda}\right)^2 \right)} \right],
~~\lambda = \frac{a(t)}{k}~. \la{gfR}
\ee
where $\lambda_c$ is a function of $R$ and is the Compton 
wavelength of the scalaron \cite{S07}. Eq. \eqref{gfR} is the equivalent form 
of (we set $a_0=1$) \cite{T07}
\be 
\frac{G_{\rm eff}(z,k)}{G}=\left(\frac{d f}{dR} \right)^{-1} 
\left[ \frac{1+4 \left(\frac{ d^2 f}{d R^2}/ \frac{ d f}{d R}\right) \cdot k^2 \, 
(1+z)^2}{1+3 \left(\frac{ d^2 f}{d R^2}/ \frac{ d f}{d R}\right) \cdot k^2 \, (1+z)^2} \right] \la{g2fr}
\ee
with $\lambda^2_c(R) =3 \frac{ d^2 f}{d R^2}/{\frac{ d f}{d R}}$.
In viable $f(R)$ models, all relevant cosmic scales satisfy 
$\lambda \gg \lambda_c(R)$, with $\frac{df}{dR}=1$ to high accuracy, deep in the matter era. 
Hence the standard growth of perturbations is regained during that era. 

Now, as $\frac{d^2f}{dR^2}>0$ \cite{S07} (which is a crucial assumption for the avoidance of 
ghost instabilities), the factor in front of the brackets in 
\eqref{gfR} increases when $R$ decreases with the expansion, and thus it is always larger than one. 
The expression inside the brackets in \eqref{gfR} is obviously always larger than one too.
So we have for $f(R)$ models that $G_{\rm eff}>G$ for any scale at any time.

At low redshifts further, as the critical length $\lambda_c$ increases significantly with the 
decrease of matter density and of the Ricci scalar $R$, the expression inside the brackets 
can become as large as $\frac{4}{3}$ in the present era on scales $\lambda\ll \lambda_c$. 
Hence the growth of matter perturbations on these scales will be enhanced compared to the 
standard growth.
Note that this does not exclude the possibility for $G_{\rm eff}(z)$ to evolve non monotonically 
as a function of $z$. Indeed, $G_{\rm eff}(z)$ can, and generically does, increase with $z$ on 
some interval in the present era, however always satisfying $G_{\rm eff}(z)>G$.
Note that \eqref{gfR} uses also $\frac{df}{dR}>0$, besides $\frac{d^2f}{dR^2}>0$, ensuring 
the absence of ghost. 

It is important to emphasize that the result presented above, \ie $\frac{G_{\rm eff}(z)}{G}>1$, 
is independent of the background expansion in contrast to the results we will derive in the 
next subsection in scalar-tensor gravity models.

\subsection{(Massless) Scalar-Tensor Gravity}
The action for this family of scalar-tensor (ST) gravity models reads 
(see e.g. \cite{EspositoFarese:2000ij})
\be
{\cal S}= \int d^4 x \sqrt{-g} \left[ \frac{1}{2}F(\phi)R - \frac{1}{2}Z(\phi) 
g^{\mu\nu} \partial_\mu \phi \partial_\nu \phi - U(\phi) \right] + S_m \la{eq:actionSctens}
\ee
where $R$ is the Ricci scalar and $S_m$ is the matter action which does not involve 
the scalar field $\phi$. Note that the coupling of these matter 
components to gravity is the same as in GR.
In what follows we set $Z(\phi) =1$ and we consider $U>0$. This means that we are dealing 
with situations where the Brans-Dicke coefficient $\omega_{BD}$ is positive (see below). 

We consider the flat Friedmann Lemaître Robertson Walker metric (FLRW), which is given by
\be
ds^2 = -dt^2 + a^2(t) \left[dr^2 + r^2 (d\theta^2 + \sin^2\theta \ d\phi^2) \right] 
\la{eq:FRWmetric}.
\ee
Then it is straightforward to show that the dynamical equations of the system are
\ba
3F H^2 &=&  \rho_m +{1\over 2} \dot\phi^2 - 3 H \dot F + U \la{eq:eqmotion1}\\
-2F \dot H  &=& \rho_m + \dot \phi^2 +\ddot F - H \dot F  \la{eq:eqmotion2}
\ea
In Eq. \eqref{eq:eqmotion1} and Eq. \eqref{eq:eqmotion2}, a homogeneous scalar field and a 
homogeneous comoving perfect dustlike fluid are assumed $(p_m=0)$. 

Setting $h(z)\equiv  \frac{H(z)}{H_0}$, $u\equiv \frac{U}{U_0}$ and 
$\Omega_{U,0}\equiv \frac{U_0}{3 F_0 H_0^2}$, 
Eq. \eqref{eq:eqmotion1} and Eq. \eqref{eq:eqmotion2} can be rewritten in terms of the  
redshift $z$ as follows
\begin{widetext}
\ba 
F'' &+& \left[ (\ln h)' - \frac{4}{1+z} \right] F' + 
\left[\frac{6}{(1+z)^2} - \frac{2}{(1+z)}(\ln h)' \right] F = 
\frac{6 u}{(1+z)^2 h^2}F_0~\Omega_{U,0} + 3 (1+z) h^{-2} F_0 \Omega_{m,0} \la{eq:eqmotionz1} \\ 
\frac{\phi'^2}{6} &=& -\frac{F'}{1+z} + \frac{F}{(1+z)^2} 
-\frac{F_0 u}{(1+z)^2 h^2} \Omega_{U,0} - \frac{F_0(1+z)}{h^2} \Omega_{m,0}~, \la{eq:eqmotionz2} 
\ea
\end{widetext}
where a prime stands for a derivative with respect to $z$. 
The first equation is a second order master equation for the quantity $F$ which is obtained 
by eliminating the kinetic term of the scalar field $\phi$. The second equation is an  
algebraic equation for the scalar field kinetic term once the equation for $F$ is solved. 
For our purposes however, we want rather to eliminate the potential energy $U$ and combining 
\eqref{eq:eqmotionz1}, \eqref{eq:eqmotionz2} we easily get the following equation
\begin{widetext}
\be
\phi'^2 = - F'' - \left[ \left( \ln h \right)'+\frac{2}{1+z} \right] F'+ 
2 \frac{ \left(\ln h \right)'}{1+z} F - 3 \Omega_{m,0} (1+z) F_0 h^{-2} \la{eq:phiz}
\ee
\end{widetext}
For our later calculations, it is convenient to introduce the quantity $\Delta^2$.  Its value 
today relative to $F_0$ is
\be
\Delta^2 \equiv \frac{\phi'^2_0}{F_0} = 6\left( \Omega_{DE,0}-\Omega_{U,0} - \frac{F'_0}{F_0} 
                                     \right)~,   \la{Delta}
\ee
where $\Omega_{DE,0}=1-\Omega_{m,0}$. The last equality follows from the Friedmann equations 
and in particular from Eq. \eqref{eq:eqmotion2}. In our universe, $\Delta^2$ is 
a positive quantity so that our notation is not confusing. 
Indeed, the right-hand side of \eqref{eq:phiz} is positive whenever the 
representation with $Z=1$ applies ($\omega_{BD}>0$). This is the case in our universe today 
and on very low redshifts. 
Once the background is fixed, \eqref{eq:phiz} expresses the kinetic term of the scalar field 
in terms of $F$ and its derivatives. 
We return now to the quantity $G_{\rm eff}$ on which we want to focus. In terms of the 
redshift $z$, $G_{\rm eff}$ can be written as
\ba
G_{\rm eff} &=& G_N \frac{2F + 4\frac{F'^2}{\phi'^2}}
         {2F + 3 \frac{F'^2}{\phi'^2}} \\  \la{eq:geffz}
       &=& G_N \left( 1 + \frac{1}{3 + 2 \omega_{BD}} \right)~, \la{geffomBD}
\ea
where we have introduced the Brans-Dicke parameter 
\be
\omega_{BD} = F \frac{\phi'^2}{F'^2}~, \la{omegaBD} 
\ee
and we have set 
\be
G_N\equiv \frac{1}{8\pi F}~.\la{GN} 
\ee

Notice that we have considered the massless scalar-tensor gravity model. This 
means physically 
that no screening (chameleon) mechanism is at work here, in contrast to the $f(R)$ models 
considered in the previous subsection. In the $f(R)$ models the mass term is central in the 
chameleon mechanism where locally $R$ and the mass become very large which enables the model 
to evade all local constraints. This is not so for our massless ST model, in particular this 
is why we have $G_{{\rm eff},0}=G$ in this case. 

Solar system constraints imply {\it today} the very strong 
inequality \cite{GPRS06}
\be
\omega_{BD,0} = \frac{\Delta^2}{\left( F'_0/F_0 \right)^2} > 4 \times 10^4~, \la{ssconstr} 
\ee
hence we have in particular 
\be
G_{{\rm eff},0} = G = G_{N,0}~,\la{G0}
\ee
where $G$ is the usual Newton's constant. We see that $\Delta^2$ is positive 
as said above.
Let us consider now the evolution of $G_{\rm eff}$. On low redshifts, we can write 
the Taylor expansion 
\be 
G_{\rm eff}(z) = G_{\rm eff,0} + G'_{\rm eff,0}~z + G''_{\rm eff,0}~\frac{z^2}{2} + \ldots 
\la{eq:geffzexpans}
\ee
The systematic expansion at low redshifts of all basic physical quantities in this ST gravity 
model was performed earlier \cite{GPRS06} (see also Ref.\cite{Nesseris:2006hp}). Here we 
extend these results by considering their 
implication for the low $z$ expansion of the effective gravitational constant $G_{\rm eff}$ up 
to second order (the first order was already derived there).

Before proceeding with the calculation of the coefficients in the expansion 
\eqref{eq:geffzexpans} we return to the consequences of solar system constraints.  
We have the following expression for $\omega_{BD,0}$ \cite{GPRS06}
\be
\omega_{BD,0} = \frac{6( \Omega_{DE,0}-\Omega_{U,0} - \frac{F'_0}{F_0} )}{\frac{F'^2_0}{F^2_0}}
\la{omegaBD0}
\ee
As we have said above, see \eqref{ssconstr}, $\omega_{BD,0}$ is a very large quantity.
Hence solar system constraints imply 
\be
\left|\frac{F'_0}{F_0}\right| \lesssim 10^{-2}~. \la{F1}
\ee
This strong inequality will considerably simplify all calculations and 
will be assumed everywhere below. 
Using \eqref{geffomBD}, \eqref{omegaBD} and \eqref{G0} the following 
results are obtained straightforwardly  
\begin{widetext}
\ba
G'_{{\rm eff}} &=& - \frac{F'}{8\pi F^2} \left( 1 + \frac{1}{3 + 2 \omega_{BD}} \right) 
              + \frac{1}{8\pi F} \left(-\frac{2 \omega_{BD}' }{(3 + 2 \omega_{BD})^2} \right)\\
G_{{\rm eff}}'' &=& 2 \frac{F'^2}{8\pi F^3}  \left( 1 + \frac{1}{3 + 2 \omega_{BD}} \right) 
     - \frac{F''}{8\pi F^2}  \left( 1 + \frac{1}{3 + 2 \omega_{BD}} \right)\\
     &+& 4 \frac{F'}{8\pi F^2}\left( \frac{\omega_{BD}' }{(3 + 2 \omega_{BD})^2} \right) 
     + \frac{1}{8\pi F} \left(- \frac{2 \omega_{BD}'' }{(3 + 2 \omega_{BD})^2}  
       + \frac{8 \omega_{BD}'^2 }{(3 + 2 \omega_{BD})^3}~.   \right)
\ea
\end{widetext}
After some straightforward calculation, using \eqref{F1}, we finally obtain to 
leading order 
\ba
G'_{{\rm eff},0} &\sim& -\frac{F'_0}{F_0}~G \ll G\\
G''_{{\rm eff},0} &\simeq& \left[ -\frac{F''}{F} + \frac{F''^2}{F~\phi'^2}  \right]_0~G
\ea
Note that the leading order of $G'_{{\rm eff},0}$ is proportional to $\frac{F'_0}{F_0}$ in 
agreement with the result obtained in \cite{GPRS06}. 
The expression for $G''_{{\rm eff},0}$ can be further simplified using \eqref{eq:phiz} 
at $z=0$ which takes the  form
\be
\phi_0'^2 = - F_0'' + 2 \left(\ln h \right)_0'~ F_0 - 3 \Omega_{0,m} F_0 \la{eq:phiz0}
\ee
When substituted in \eqref{eq:geffzexpans}, we obtain
\begin{widetext}
\ba
G_{\rm eff}(z) \simeq  &G& \left( 1 + \frac{F''_0}{F_0} \left[-2 + \Delta^{-2} 
  \left( \left(\ln h^2 \right)_0' - 3 \Omega_{m,0}\right) \right] \frac{z^2}{2} \right)\\
\simeq &G& \left( 1 + \frac{F''_0}{F_0} \left[-1 + \frac{3}{2} \Delta^{-2} 
  (1 + w_{DE,0})(1-\Omega_{m,0}) \right] z^2 \right)~.\la{Geffz2}
\ea  
Hence the variation of $G_{\rm eff}$ on low redshifts, and in particular its departure from $G$, 
depends crucially on the magnitude and on the sign of $\frac{F''_0}{F_0}$.
For $\frac{F'_0}{F_0}\ll 1$, we have 
\be
\frac{F''_0}{F_0} = 3(w_{DE,0} + 1)\Omega_{DE,0} - 6 (\Omega_{DE,0} - \Omega_{U,0})~. \la{F2ss}
\ee
When this is substituted in \eqref{Geffz2}, we finally obtain 
\be
G_{\rm eff}(z) \simeq G \left( 1 + 
\Big[ 3(w_{DE,0} + 1)\Omega_{DE,0} - 6 (\Omega_{DE,0} - \Omega_{U,0}) \Big]
\left[-1 + \frac{3}{2} \Delta^{-2} (1 + w_{DE,0})\Omega_{DE,0}) \right] z^2 \right)~.\la{Geffw0}
\ee
\end{widetext}
Before proceeding with our investigation, a first important remark is that 
\eqref{Geffw0} simplifies considerably for a $\Lambda$CDM background 
to yield
\ba
G_{\rm eff}(z) \simeq &G& \left( 1 - \frac{F''_0}{F_0} ~z^2 \right) \nn \\
\simeq  &G& \left( 1 + 6 (\Omega_{DE,0} - \Omega_{U,0})~z^2 \right) \la{GeffL}
\ea
Two cases can arise depending on the sign of $\Omega_{DE,0} - \Omega_{U,0}$. 

a) The most natural case to consider is  
\be
\Omega_{DE,0} - \Omega_{U,0}\gg \left|\frac{F'_0}{F_0}\right|~, \la{Omdif}
\ee 
while the solar system constraint \eqref{ssconstr} is satisfied using 
\eqref{omegaBD0}, \eqref{F1}. 
In this case we have from \eqref{Delta}
\be
\Delta^2 \approx  6 (\Omega_{DE,0} - \Omega_{U,0})~, \la{Deltaapprox}
\ee
and we obtain for a $\Lambda$CDM background from \eqref{F2ss}, \eqref{GeffL}  
\be 
\frac{F''_0}{F_0} \simeq - 6(\Omega_{DE,0} - \Omega_{U,0})\approx -\Delta^2 < 0~,
\ee
and 
\be
G_{\rm eff}(z) = G \left( 1 + 6 (\Omega_{DE,0} - \Omega_{U,0})~z^2 \right) \approx 
G \left( 1 + \Delta^2 z^2 \right) \la{GeffzaL}
\ee
which is a central result of our calculation. Hence for a $\Lambda$CDM background, 
$G_{\rm eff}(z)$ will increase rather than decrease in the past on low redshifts. 
It is seen from \eqref{F2ss} that this result applies whenever dark energy (DE) is of the phantom 
type today and satisfies $w_{DE,0}<-1$. 
It can even hold for some small range of values satisfying $w_{DE,0}\gtrsim -1$.

\begin{figure}
\begin{centering}
\includegraphics[scale=0.48]{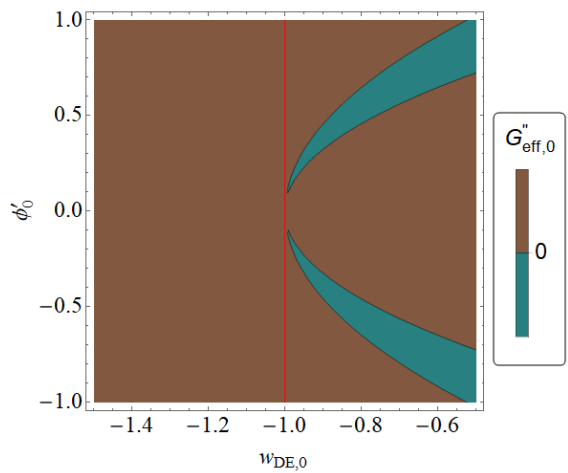}
\par\end{centering}
\caption{The second derivative of $G_{\rm eff}$ in the parametric space  ($\phi_0'-w_{DE,0}$) for $\Omega_{m,0}=0.3$. The blue regions denote the areas where $G_{{\rm eff},0}''<0$ while the brown regions correspond to $G_{{\rm eff},0}''>0$}  
\label{fig:Sctenswcdm}
\end{figure}

It is possible however to get a decreasing $G_{\rm eff}(z)$ if we move away from $\Lambda$CDM 
towards higher values of $w_{DE,0}$ satisfying $w_{DE,0}>-1$. By inspection of \eqref{Geffw0}, 
this is the case if the following inequality holds
\be
\Delta^2 < 3(w_{DE,0} + 1)\Omega_{DE,0} < 2 \Delta^2 \la{ineqw0}~.
\ee
The inequality \eqref{ineqw0} can be easily satisfied for a large number of parameter 
values as we show on Figure \ref{fig:Sctenswcdm}. We conclude that, for case a), 
$w_{DE,0}>-1$ is necessary in order to have a decreasing $G_{\rm eff}(z)$ on low redshifts.   

b) In principle, there is also the possibility $\Omega_{DE,0} - \Omega_{U,0} < 0$. In 
that case however we have from \eqref{omegaBD0}
\be
\left| \Omega_{DE,0}-\Omega_{U,0} \right| < 
                           \left| \frac{F'_0}{F_0}\right| \lesssim 10^{-4} ~,
\ee
and also
\be 
\Delta^2 < \left| \frac{F'_0}{F_0}\right| \lesssim 10^{-4}~.
\ee
In \eqref{F2ss} we should now discard the second term on the right hand side as we have 
done for all terms proportional to $\frac{F'_0}{F_0}$ and we simply write
\be
\frac{F''_0}{F_0} = 3(w_{DE,0} + 1)\Omega_{DE,0}~, \la{F2Lb}
\ee
and \eqref{Geffw0} becomes 
\begin{widetext}
\be
G_{\rm eff}(z) \simeq G \left( 1 + 
\Big[ 3(w_{DE,0} + 1)\Omega_{DE,0} ) \Big] \left[-1 + 
   \frac{3}{2} \Delta^{-2} (1 + w_{DE,0})\Omega_{DE,0}) \right] z^2 \right)~.\la{Geffw0b}
\ee
\end{widetext}
Again, let us consider first a \lcdm background. 
In that case  $G''_{{\rm eff},0}$ is of the same magnitude as, 
or even smaller than, $G'_{{\rm eff},0}$ and the corresponding $G_{\rm eff}(z)$ is essentially 
constant on low redshifts as we see immediately from \eqref{Geffw0b}. 

Moving away from $w_{DE,0}=-1$, we obtain again as in the previous case, an increasing 
$G_{\rm eff}(z)$ in the past for $w_{DE,0}<-1$ \emph{and} for $w_{DE,0}>-1$ whenever 
$1+w_{DE,0}\gg \Delta^2$. An essentially constant $G_{\rm eff}(z)$ is obtained for 
$1+w_{DE,0}\approx \Delta^2$, in other words for $1+w_{DE,0}$ vanishingly small.  
Hence, case b) does not lead to a decreasing $G_{\rm eff}(z)$ on low $z$. 

To summarize all possibilities, for $w_{DE,0}\le -1$, either $G_{\rm eff}(z)$ increases with 
$z$ on small redshifts, or else it is essentially constant. In contrast, for 
$w_{DE,0}> -1$ configurations are easily found (case a)) that yield a decreasing 
$G_{\rm eff}(z)$. Note that solar system constraints play an essential role in these 
derivations.

We illustrate these results with Fig. \ref{fig:Sctenswcdm}.
Clearly, $G_{{\rm eff},0}''<0$ (blue regions) can only be achieved for $w>-1$. This behavior 
remains valid for different values of $\omm$.  The results presented in this section 
assume that $G'_{{\rm eff},0}\simeq 0$ (or equivalently $F_0'\simeq 0$) due to solar system 
constraints. 
In the presence of screening this assumption may not be necessary as in that case the 
cosmological behavior of $G_{\rm eff}$ gets decoupled from the corresponding behaviour in the 
solar system where the mean curvature and density are significantly larger than in 
cosmological scales. However, as we have seen with $f(R)$ models, this does not necessarily 
imply that a decreasing $G_{\rm eff}(z)$ is allowed and actually in these models, it is not allowed. 
 
\section{The growth index $\gamma$}
\label{sec:gamma}
In this Section, we will not assume any specific massless ST model, but rather consider a parametrization of $G_{\rm eff}$ consistent with $G_{\rm eff,0}=G$ at $z=0$ and deep in the matter era. 
In \cite{Kazantzidis:2018rnb}, such a parametrization of \geffz was suggested
\ba
\frac{G_{\textrm{eff}}}{G} &=& 1+g_a(1-a)^n - g_a(1-a)^{n+m} \nn \\
&=&1+g_a\left(\frac{z}{1+z}\right)^n - g_a\left(\frac{z}{1+z}\right)^{n+m}~, \la{eq:geffansatz}
\ea
where $n\geq 2$ and $m>0$. Throughout this Section we set $m=n$. The parametrization \eqref{eq:geffansatz} has been used in Refs 
\cite{Nesseris:2017vor,Kazantzidis:2018rnb} to show that in the context of a 
Planck15/$\Lambda$CDM background the best fit value of the parameter $g_a$ indicated by 
$\fs$ subsamples is negative and that it is inconsistent with zero or positive values at a 
level more than $3\sigma$. This result tends to indicate that a decreasing $G_{\rm eff}(z)$ is 
significantly favored by the $\fs$ data if the background expansion is close to the one given 
by the Planck15/$\Lambda$CDM parameter values. However, as we will see below using the parametrization \eqref{eq:geffansatz},  a rapidly decreasing \geffz on low redshifts is ruled out by the SNIa data for 
$n\le 5$

It is then possible to find the corresponding growth index $\gamma$ once the 
background expansion is also fixed. In other words, independently of the specific 
modified gravity model that produces \eqref{eq:geffansatz} \emph{and} the \plcdm background 
expansion, we can find the resulting growth index. 

In particular the quantity $f(z)$ obeys the following equation
\be 
\frac{df}{d\ln a} + f^2 + \frac{1}{2} \left(1 - \frac{d \ln \Omega_m}{d\ln a} \right) f = 
                              \frac{3}{2} \frac{G_{\rm eff}}{G} \Omega_m~,\la{df}
\ee
where $\Omega_m=\frac{\Omega_{m,0} ~a^{-3}}{H(a)^2/H_0^2} =\frac{\Omega_{m,0} ~a^{-3}}{h^2(a)}$ and $\delta$ can be obtained directly through
\be 
\delta(a) = \delta_i~{\rm exp} \left[ \int_{lna_i}^{lna} f d(lna') \right]~.
\ee 
The growth rate $f$ can always be written as 
\be
f = \Omega_m^{\gamma}~,\la{Omgamma}
\ee
where $\gamma$ is nearly constant in GR $\gamma \approx 0.55$ \cite{Linder:2005in}. 
For many modified theories, $\gamma$ departs from this quasi-constant behaviour \cite{Gannouji:2008wt} and 
can be written at small $z$ as $\gamma=\gamma_0+\gamma_0'~z$.
Using Eqs. \eqref{df}, \eqref{Omgamma}, we have 
\begin{align}
2 \ln \Omega_m~\frac{d\gamma}{d\ln a} + (2\gamma-1)~\frac{d\ln \Omega_m}{d\ln a} + 1 + 2\Omega_m^{\gamma} \nonumber\\
- 3 \frac{G_{\rm eff}}{G_N} \Omega_m^{1-\gamma} = 0
\la{eq:mas}
\end{align}
So if we know the background expansion and $\Omega_{m,0}$, as well as the behaviour 
of $G_{\rm eff}$, 
we can calculate $\gamma$ \cite{Gannouji:2018col}.
Assuming a \plcdm background while $G_{\rm eff}$ is of the form \eqref{eq:geffansatz} we are left with a first order differential 
equation for $\gamma$. We fix the initial condition in the past in order to find 
$\gamma(z)$, and therefore $\gamma_0\equiv \gamma(0)$ and $\gamma_0'\equiv \gamma'(0)$,
for each $(g_a,n)$ (see Fig. \ref{fig:SNIa-gamma}).
Notice that initial conditions (in the past) are essentially irrelevant at the present time 
because of the presence of an attractor so we get the same behaviour at late time.
 
For the case $n=m$ in Eq. \eqref{eq:geffansatz}, our result are consistent with previous results derived in \cite{Gannouji:2018col}: a 
weaker gravitational constant ($G_{\rm eff}<G$) implies $\gamma_0>\gamma_0^{\Lambda CDM}$ 
for a given background while a stronger gravitational constant ($G_{\rm eff}>G$) implies 
$\gamma_0<\gamma_0^{\Lambda CDM}$.

\begin{figure}
\begin{centering}
\includegraphics[scale=0.28]{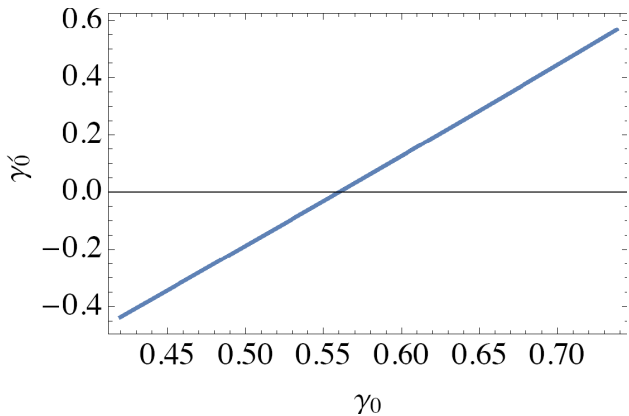}
\par\end{centering}
\caption{The (linear) relation between $\gamma'_0$ and $\gamma_0$ is shown 
for \emph{any} values of $g_a$ and $n$ and a fixed background (\plcdmnospace). This relation is 
independent of $g_a$ because we have $G_{{\rm eff},0}=G$ by construction for all parameter values $g_a$ and $n$.}
\label{fig:gamma1_0}
\end{figure}

\begin{figure}[H]
\centering
\includegraphics[scale=0.65]{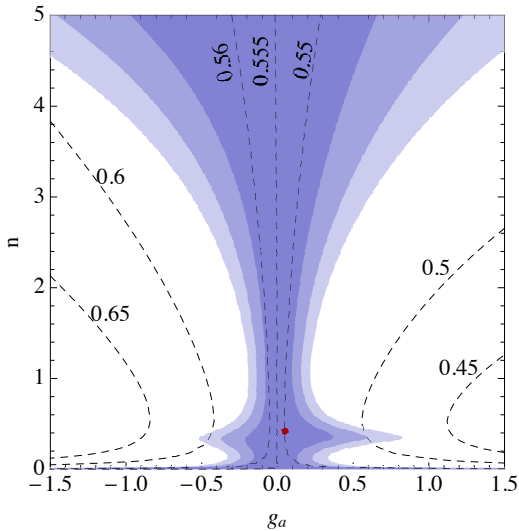}
\caption{Constraints at 1$\sigma$, 2$\sigma$, 3$\sigma$ level from the SNIa data are shown in the 
$g_a,n$ plane. The dashed curves correspond to couples with the same value $\gamma_0$. 
The corresponding value of $\gamma'_0$ is easily obtained from Fig. \ref{fig:gamma1_0}.}
\label{fig:SNIa-gamma}
\end{figure}

Also in accordance with \cite{Polarski:2007rr}, we found that $\gamma_0'$ is 
linearly related to $\gamma_0$ for different values of the free parameters of the model, see Fig. \ref{fig:gamma1_0}. In fact, considering Eq. \eqref{eq:mas} at $z=0$, we have
\begin{widetext}
\be
\gamma'_0 = \frac{1}{2 \ln \Omega_{m,0}} 
\left[ (2\gamma_0 -1)~3 w_{DE,0} (1-\Omega_{m,0}) + 1 + \\
+ 2 \Omega^{\gamma_0}_{m,0} 
   - 3~\frac{G_{{\rm eff},0}}{G} \Omega^{1-\gamma_0}_{m,0} \right]~.
\la{dgamma0b}
\ee
\end{widetext}

In our case we have by construction $\frac{G_{{\rm eff},0}}{G}=1$ for all parameters $n, m, g_a$. 
Hence the relation $\gamma'_0 =f(\gamma_0)$ from (\ref{dgamma0b}) is the same as shown 
in Fig. \ref{fig:gamma1_0}

We have also considered constraints from SNIa data and we find that these do not 
significantly favor $g_a<0$ (see Fig. \ref{fig:SNIa-gamma}). The distance modulus for the SNIa data can be written as \cite{Arnett:1982}
\begin{align}
\mu = \mu_{\Lambda CDM} + \frac{15}{4} \log~\frac{G_{\rm eff}}{G_{{\rm eff},0}}~,
\end{align}
where the additional term comes from the modification of the luminosity distance as a result 
of modified gravity. In our analysis we use the latest Pantheon Sample \cite{Scolnic:2017caz} of 1048 
SNIa ranging from $0.01 < z < 2.3$.

Clearly the SNIa data are not consistent with $g_a<-0.3$ at the $3 \sigma$ level for $n=2$. For higher values of $n$ however, significantly lower values of $g_a$ are allowed. Similar results were obtained for the CMB data (ISW effect) in Ref. \cite{Nesseris:2017vor}. These results indicate that the tension of the growth data with \plcdm can only be partially physical. At least part of this tension is probably due to statistical and/or systematic effects of the growth data. However this tension points to a mildly decreasing \geffz rather than to an increasing, or even a constant, \geffznospace.

To complete this section, we provide the values of ($\gamma_0,\gamma'_0$) 
corresponding to parameters $(g_a,n)$ favored by the $f\sigma_8$ data (see Table \ref{table:gamma}). For each $n$, the best value of $g_a$ and therefore $G_{eff}(z)$ was 
obtained in \cite{Nesseris:2017vor} as shown in Table \ref{table:gamma}. 
\begin{table}[H]
\centering
\begin{tabular}{|c|c|c|c|}
\hline 
$n$ & $g_{a}$ & $\gamma_{0}$ & $\gamma_0'$\tabularnewline
\hline 
\hline 
0.343 & -1.200 & 0.686 & 0.398\tabularnewline
\hline 
2 & -1.156 & 0.629 & 0.219\tabularnewline
\hline 
3 & -1.534 & 0.620 & 0.189\tabularnewline
\hline 
4 & -2.006 & 0.615 & 0.174\tabularnewline
\hline 
5 & -2.542 & 0.612 & 0.165\tabularnewline
\hline 
6 & -3.110 & 0.611 & 0.160\tabularnewline
\hline 
\end{tabular}
\caption{Corresponding values of ($\gamma_0,\gamma'_0$) for various $(n,g_a)$ 
favored by $f\sigma_8$ data alone. The behaviour of $\gamma$ is a characteristic signature for a decreasing 
$G_{\rm eff}<G$ on low redshifts ($G_{{\rm eff},0}=G$). We remind that all values $n\le 5$ are ruled out by SNIa data.}
\label{table:gamma}
\end{table}

Finally, as we have stressed earlier, $f(R)$ models always satisfy $G_{\rm eff}>G$. 
Therefore, for all background evolutions that would produce $\gamma_0\approx 0.55$ inside 
GR, the value of $\gamma_0$ obtained in $f(R)$ models will satisfy $\gamma_0 \lesssim 0.55$ in 
accordance with \cite{Tsujikawa:2009ku}.


\section{Summary and discussion} 
\label{sec:Conclusion}
A $G_{\rm eff}(z)<G$ at low redshifts could alleviate the tension between \plcdm and the growth 
data $\fs$. In this work we have studied the implications of such a \geffz for two classes of  
modified gravity DE models. 

The $f(R)$ DE models cannot produce such a behaviour. More generally they cannot allow for 
$G_{\rm eff}(z)<G$ irrespective of the background expansion \cite{Polarski:2016ieb}. 
We have further shown that in (massless) scalar tensor theories, a decreasing \geffz 
at low redshifts is not possible for a \lcdm background. 
However this behaviour is possible if we consider $w_{DE,0}>-1$, and a substantial decrease 
of $G_{\rm eff}(z)$ requires a substantial departure from $w_{DE,0}=-1$.  

We have further shown that any model with the required behaviour of $G_{\rm eff}(z)$ in a 
\lcdm background will exhibit a characteristic signature of its growth index $\gamma$, with 
$0.61\lesssim \gamma_0\lesssim 0.69$ and a non-negligible slope $\gamma_1$ at $z=0$, 
$0.16\lesssim  \gamma_1\lesssim 0.4$. Once redshift space distortion data become more 
accurate, it will be possible not only to discriminate between these models and \lcdmnospace, but 
also to confirm or to rule out the decreasing \geffz which is required to explain the data. 

While it is known that some modified gravity DE models can have $G_{\rm eff}(z)<G$ in 
principle \cite{Linder:2018jil}, 
it is interesting that two prominent representatives of viable modified gravity DE models 
cannot produce such a behaviour. If this behaviour plays a role in the solution 
to the existing tension in the data between \plcdm and the redshift space distortion data, 
our results imply that more elaborate modified gravity models are required.

\section*{Acknowledgements}

The work of R. Gannouji is supported by Fondecyt project No 1171384. 
D.P. aknowledges KASI (Daejeon, Korea) for hospitality. 

\raggedleft
\bibliography{bibliography}

\end{document}